\documentclass[conference]{IEEEtran}
\IEEEoverridecommandlockouts
\usepackage{cite}
\usepackage{amsmath,amssymb,amsfonts}
\usepackage{graphicx}
\usepackage{textcomp}
\usepackage{xcolor}
\usepackage{booktabs}
\usepackage{array}
\usepackage{url}

\urldef{\PowerAgentBenchURL}\url{https://github.com/Power-Agent/PowerAgentBench}

\def\BibTeX{{\rm B\kern-.05em{\sc i\kern-.025em b}\kern-.08em
    T\kern-.1667em\lower.7ex\hbox{E}\kern-.125emX}}

\begin{document}

\title{PowerAgentBench-SS: A Benchmark for Agentic AI in Power System Steady-State Studies%
\thanks{

Corresponding author: Costas Mylonas (kmylonas@ubitech.eu). 

The code repository of this paper is maintained by PowerAgent Community at: \PowerAgentBenchURL.}
}

\author{ 
\IEEEauthorblockN{
Costas Mylonas\IEEEauthorrefmark{1}\IEEEauthorrefmark{5},
Magda Foti\IEEEauthorrefmark{1},
Andrea Pomarico\IEEEauthorrefmark{2},
Matheus Duarte\IEEEauthorrefmark{3},
Qian Zhang\IEEEauthorrefmark{4},
and Emmanouel Varvarigos\IEEEauthorrefmark{5}
} 

\IEEEauthorblockA{\IEEEauthorrefmark{1}Energy Digitalization Group, UBITECH, Athens, Greece}
\IEEEauthorblockA{\IEEEauthorrefmark{2}Department of Energy, Politecnico di Milano, Milan, Italy}
\IEEEauthorblockA{\IEEEauthorrefmark{3}EnliteAI, Vienna, Austria}
\IEEEauthorblockA{\IEEEauthorrefmark{4}School of Engineering and Applied Sciences, Harvard University, Allston, MA, USA}
\IEEEauthorblockA{\IEEEauthorrefmark{5}Department of Electrical and Computer Engineering, National Technical University of Athens, Athens, Greece}
}

\maketitle

\begin{abstract}
Power system benchmarks usually evaluate numerical solvers, prediction models, or sequential controllers. These benchmarks are necessary, but they do not directly test whether a Large Language Model (LLM) agent can execute an engineering workflow: inspect a grid case, select tools, call simulators, screen contingencies, propose admissible mitigations, validate results, and produce an auditable evidence trail. This paper introduces PowerAgentBench-SS, a steady-state benchmark framework for evaluating tool-using agents in power system operation and planning studies. The benchmark exposes public case data, action constraints, a tool API, and a validation budget to an agent, while a hidden evaluator recomputes physical validity and scores the submitted report. We define the agent interface, tool contract, evidence log, and risk-sensitive metrics, including submitted recall, evidence-backed recall, found recall, false-safe penalties, severity regret, residual violation score, action cost, tool-use efficiency, and workflow diagnostics. To make the framework concrete, we instantiate the protocol in a reproducible DC thermal $N$-\(2\) contingency-search pilot on deterministic IEEE 39-bus operating-point variants, with scripted baselines, an LLM JSON-command adapter, three locally hosted Ollama LLM agents, and one OpenAI API agent. The results show why solver-only or answer-only evaluation is insufficient: agents are distinguished not only by top-contingency discovery, but also by validation-budget use, explicit submission, type coercions, duplicate validations, evidence-backed reporting, and mitigation behavior.
\end{abstract}


\section{Introduction}
Reliable power system operation and planning are not single-step prediction tasks. Engineers build cases, inspect data, run power-flow studies, screen contingencies, identify thermal or voltage violations, design mitigations, validate corrected cases, and write reports that another engineer can audit. This motivates an evaluation target different from standard Optimal Power Flow (OPF) optimality, classifier accuracy, or direct question answering: can an AI system behave as an engineering agent that uses tools and produces physically verified evidence?

This question is becoming urgent as LLM-based agents move from text generation toward tool use. General agent benchmarks already emphasize interactive environments, external tools, hidden execution-based scoring, and repeated-trial reliability. AgentBench evaluates LLMs in multi-turn environments \cite{liu2024agentbench}. SWE-bench uses real software issues and test-based validation \cite{jimenez2023swebench}. WebArena provides reproducible web environments with functional task completion \cite{zhou2023webarena}. OSWorld evaluates agents in real computer environments with task setup and execution-based checks \cite{xie2024osworld}. $\tau$-bench studies domain APIs, policies, and pass@$k$ reliability \cite{yao2024taubench}. MLAgentBench evaluates iterative machine learning experimentation \cite{huang2023mlagentbench}. Power systems require the same agentic ingredients, but with an additional safety constraint: an agent must not declare an operating point secure, dismiss a contingency as safe, or certify a mitigation plan unless that claim is supported by validated power system calculations. We refer to such unsupported security claims for actually unsafe cases as false-safe conclusions.

AI for power systems is also moving toward agentic workflows. The PowerAgent vision combines domain foundation models, standardized tool interfaces, reusable workflows, and human supervision \cite{zhang2025poweragent}. Studies of LLMs in electric energy tasks typically emphasize tool integration, domain-specific evaluation, retrieval, and safety guardrails \cite{majumder2024llm,chen2025x,chaturvedi2025grid}. Existing power system evaluation resources provide important foundations. PGLib-OPF supports AC OPF benchmarking \cite{babaeinejadsarookolaee2019pglib}, and the ARPA-E Grid Optimization Competition evaluates large-scale security-constrained AC OPF algorithms \cite{aravena2022go}. In parallel, the mature SCOPF literature shows that practical security analysis combines filtering, approximations, decomposition, warm starts, and validation \cite{capitanescu2011scopf}. However, these resources primarily evaluate algorithms or solver pipelines, not whether an autonomous agent chooses appropriate tools, uses a limited validation budget, and gives an auditable recommendation. 

PowerAgentBench-SS addresses the above gap for steady-state studies with a focus on $N$-\(1\) and $N$-\(k\) contingency analysis, where tasks combine physical validation, combinatorial search, tool selection, and risk-sensitive scoring. Existing $N$-\(k\) studies formulate severe multiple-contingency discovery as vulnerability analysis, interdiction, and screening \cite{bienstock2009nk,pinar2010optimization,donde2008severe}, while fast $N$-\(2\) screening shows the need for intelligent pruning rather than brute-force enumeration \cite{kaplunovich2016fast}. Recent learned OPF and screening methods, such as DeepOPF, CANOS, OPFData, and reliable neural contingency screening, can accelerate this workflow \cite{pan2019deepopf,piloto2024canos,lovett2024opfdata,christianson2024reliable}. In our framework, such methods are tools available to an agent and the final claims are still checked by hidden physical validation. A complementary benchmark for power system dynamic studies is available as PowerAgentBench-Dyn \cite{zhang2026poweragentbenchdyn}.

The contributions of the current work are: (i) a user-facing benchmark protocol for testing LLM agents on steady-state power system tasks, (ii) an explicit public tool API and evidence log that separate the agent-visible environment from hidden scoring, (iii) risk-sensitive metrics and workflow diagnostics that distinguish submitted answers, evidence-backed answers, found contingencies, and mitigation outcomes, and (iv) a reproducible DC thermal $N$-\(2\) pilot on IEEE 39-bus operating-point variants with scripted baselines, locally hosted Ollama LLM agents, and an OpenAI API agent.

\section{Benchmark Design for Agents}
\subsection{Agentic Evaluation Principles}
\begin{figure}[ht]
\includegraphics[width=0.98\columnwidth]{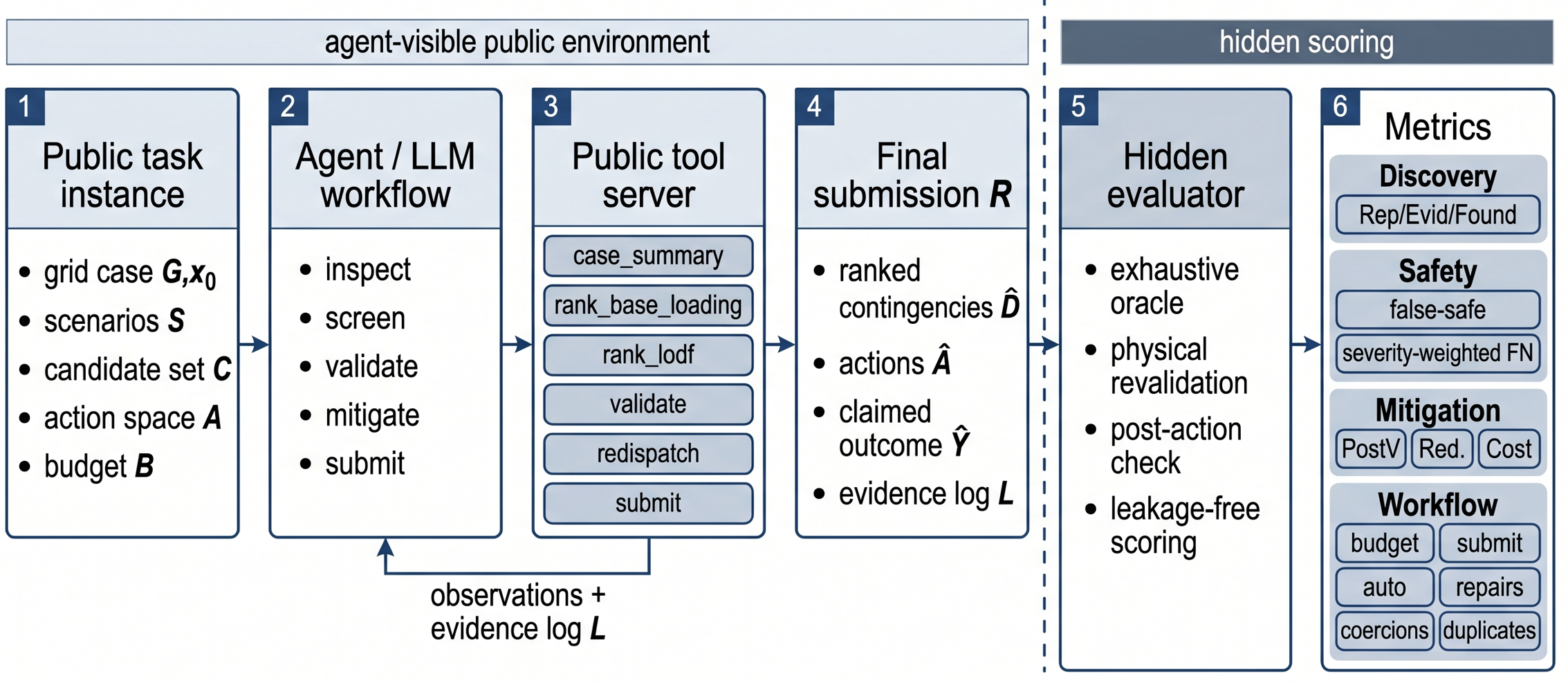}
\caption{PowerAgentBench-SS evaluation loop.}
\label{fig:framework}
\end{figure}

PowerAgentBench-SS evaluates an engineering workflow rather than a text-only answer. Fig.~\ref{fig:framework} summarizes the benchmark loop. The agent-visible environment contains the public task instance, the LLM or scripted agent, the public tool server, and the final submission. The hidden scoring side recomputes oracle rankings, physical validity, post-action violations, and workflow diagnostics after the agent submits.

The design follows five principles. First, the task must be executable: the agent interacts with analysis tools, and the benchmark checks the resulting physical state rather than grading text alone. Second, the public environment must be bounded and reproducible: the agent receives a defined case, scenario, action space, tool API, and budget. Third, scoring must be hidden: oracle labels, held-out severities, and final revalidation are not exposed during the run. Fourth, engineering rules must be explicit: admissible actions, operating limits, costs, and approval gates are part of the task definition. Fifth, reliability must be measured over multiple instances or trials, because one successful run is not enough for critical-infrastructure settings. This public/hidden separation limits benchmark leakage and helps ensure that scores reflect tool-supported engineering decisions rather than access to oracle labels or evaluator logic.

\subsection{Instance and Submission Format}
A PowerAgentBench-SS task instance is
\begin{equation}
\mathcal{I}=\left(G,x_0,\mathcal{S},\mathcal{T},\mathcal{A},B,\mathcal{E}\right).
\end{equation}
Here, \(G\) includes the static grid data: buses, branches, generators, loads, equipment parameters, and operating limits. The starting point \(x_0\) includes the public operating data for the study, such as dispatch, demand or injection profiles, and any base-case solved quantities, such as voltages or flows when applicable. The set \(\mathcal{S}\) specifies the study cases for the task depending on the benchmark level, such as the base case, N-\(1\) outages, N-\(k\) contingency candidates, operating-point scenarios, transfer-level cases, or planning variants such as load growth, retirements, resource additions, and candidate upgrades. The tool set \(\mathcal{T}\) specifies the public functions available to the agent, such as AC/DC power-flow validation, OPF or SCOPF routines, contingency screening, line-outage sensitivities, and evidence/report utilities. The action space \(\mathcal{A}\) specifies admissible controls, their physical units, bounds, step sizes, feasibility rules, and costs. The budget \(B\) specifies limits on validation calls, tool calls, turns, or runtime. Finally, \(\mathcal{E}\) implements hidden scoring by recomputing oracle results, checking evidence support, revalidating submitted actions, and scoring the final submission.

An agent returns
\begin{equation}
R=\left(\widehat{\mathcal{D}},\widehat{A},\widehat{Y},L\right),
\end{equation}
where \(\widehat{\mathcal{D}}\) is the submitted diagnosis, such as a list of violations or ranked contingencies, \(\widehat{A}\) is the proposed mitigation plan, and \(\widehat{Y}\) is the claimed post-action outcome. The log \(L\) records tool calls, arguments, observations, validation results, and the final submission. The hidden evaluator revalidates \(\widehat{\mathcal{D}}\) and \(\widehat{A}\), checks \(\widehat{Y}\), and uses \(L\) to determine which submitted cases and actions are supported by public evidence. This allows the benchmark to reward correct solutions while penalizing unvalidated, inconsistent, or unsupported claims.

PowerAgentBench-SS accepts \(R\) through three submission modes. In scripted mode, heuristic baselines and conventional algorithms are connected directly to the runner and return the same structured submission \(R\), without using a language model interface. In LLM mode, the same interaction is mediated by a JSON-command adapter, so the model never calls the evaluator directly and only observes public tool outputs. In human-supervised mode, selected actions can require approval, and the approval record is stored in \(L\). These modes let users compare heuristic baselines, tool-using LLMs, and operator-assistant workflows under the same scoring rules.

\subsection{Tool API for LLM Agents}
The tool API is the public contract between an LLM agent and the benchmark environment. For each task, it specifies the admissible tool names, argument schemas, return fields, budget effects, and error handling. In steady-state studies, the tools may screen cases, run validation, apply admissible actions, or submit the final report. The API makes tool use auditable because every tool call, argument, observation, and validation result is recorded in the evidence log \(L\).

In LLM mode, the tool API is exposed through a minimal JSON-command interface implemented by an adapter. At each turn, the prompt contains the task description, allowed tools, remaining budget, and previous tool observations and the model returns one structured command naming a tool and its arguments. The adapter is public middleware that parses the command, checks schemas and budgets, executes the corresponding public function, records the observation in \(L\), and appends that observation to the next turn.

Invalid JSON, unknown tools, invalid arguments, duplicate validations, and budget violations are recorded as workflow diagnostics. The prompt template specifies the canonical command schema and case representation. Unambiguous noncanonical outputs, such as string branch identifiers, can be repaired by the adapter, but repairs and type coercions are counted as diagnostics. 

\section{Benchmark Tasks and Evaluation}
\subsection{Steady-State Task Levels}
The task levels define power system difficulty independently of the agent implementation. Level 1 is $N$-\(1\) audit and mitigation: the agent receives a base case and a published list of single contingencies, identifies violations, selects bounded controls, validates base and contingency states, and reports evidence. Level 2 is $N$-\(k\) search and mitigation: the agent receives a candidate contingency set \(\mathcal{C}\) and order \(k\), so the study space contains \(C\subseteq\mathcal{C}\) with \(|C|=k\). Exhaustive enumeration of \(\binom{|\mathcal{C}|}{k}\) cases may be infeasible under budget, so the agent must allocate validation calls using screening, graph reasoning, learned surrogates, or adaptive search.

\subsection{Risk-Sensitive Metrics}
Let \(\mathcal{U}\) denote the hidden evaluation set scored by the evaluator, such as contingency cases or operating/planning scenarios. Let \(s(C)\ge 0\) be the hidden pre-action severity of case \(C\in\mathcal{U}\), with larger values indicating worse physical violations. Its task-specific definition may combine thermal overloads, voltage violations, infeasibility penalties, load shedding, or other engineering criteria.

Let \(\mathcal{O}_{m}\subseteq\mathcal{U}\) be the hidden oracle top-\(m\) set, i.e., the \(m\) cases with largest \(s(C)\), where \(m\) is the report size. Let \(\widehat{\mathcal{D}}\) be the submitted set of cases, and let \(\mathcal{V}\) be the set the agent validated through public tools. We report three recall metrics:
\begin{equation}
\begin{aligned}
\mathrm{R}_{\mathrm{sub}}
&=
\frac{|\widehat{\mathcal{D}}\cap \mathcal{O}_{m}|}
{|\mathcal{O}_{m}|},\\
\mathrm{R}_{\mathrm{evid}}
&=
\frac{|\widehat{\mathcal{D}}\cap\mathcal{V}\cap \mathcal{O}_{m}|}
{|\mathcal{O}_{m}|},\\
\mathrm{R}_{\mathrm{found}}
&=
\frac{|\mathcal{V}\cap \mathcal{O}_{m}|}
{|\mathcal{O}_{m}|}.
\end{aligned}
\end{equation}
\(R_{\mathrm{sub}}\) scores what the agent submits, whether or not it was validated. \(R_{\mathrm{evid}}\) gives credit only to submitted cases also present in the validation log. \(R_{\mathrm{found}}\) measures search success from \(\mathcal{V}\), even if the agent fails to convert found cases into a valid final submission.

For the submitted set, EvidenceRate measures the fraction backed by validation evidence, and UnvalidatedRate is its complement:
\begin{equation}
\begin{aligned}
\mathrm{EvidenceRate}
&=
\frac{|\widehat{\mathcal{D}}\cap\mathcal{V}|}
{\max\{1,|\widehat{\mathcal{D}}|\}},\\
\mathrm{UnvalidatedRate}
&=
1-\mathrm{EvidenceRate}.
\end{aligned}
\end{equation}
These metrics distinguish plausible-looking reports from claims supported by tool outputs.

For worst-case discovery, let \(s^\star\) be the oracle worst severity and \(s_{\mathrm{val}}\) the worst severity the agent validated:
\begin{equation}
s^\star=\max_{C\in\mathcal{U}}s(C),
\qquad
s_{\mathrm{val}}=\max_{C\in\mathcal{V}}s(C),
\end{equation}
with \(s_{\mathrm{val}}=0\) if \(\mathcal{V}\) is empty. Regret is the remaining severity gap, and Best is the normalized severity captured:
\begin{equation}
\mathrm{Regret}=s^\star-s_{\mathrm{val}},
\qquad
\mathrm{Best}=
\frac{s_{\mathrm{val}}}{\max\{\epsilon,s^\star\}},
\end{equation}
where \(\epsilon>0\) avoids division by zero. Thus, an agent can have low top-\(m\) recall but high Best if it finds the single worst case.

False-safe behavior occurs when dangerous cases are omitted or reported without validation support. Let \(\mathcal{D}\subseteq\mathcal{U}\) be the hidden dangerous set, defined by operating-limit violations, an absolute severity threshold, or a severity quantile. The dangerous set is not exposed during the run. We define reported dangerous-case precision as
\begin{equation}
\mathrm{P}_{\mathrm{danger}}=
\frac{|\widehat{\mathcal{D}}\cap\mathcal{D}|}
{\max\{1,|\widehat{\mathcal{D}}|\}}.
\end{equation}
The false-safe rate and severity-weighted false-negative score are
\begin{equation}
\mathrm{FSR}=
\frac{|\mathcal{D}\setminus(\widehat{\mathcal{D}}\cap\mathcal{V})|}
{\max\{1,|\mathcal{D}|\}},
\end{equation}
\begin{equation}
\mathrm{SWFN}=
\frac{\sum_{C\in \mathcal{D}\setminus(\widehat{\mathcal{D}}\cap\mathcal{V})}s(C)}
{\max\{\epsilon,\sum_{C\in\mathcal{D}}s(C)\}}.
\end{equation}
FSR and SWFN measure the fraction and severity-weighted share of dangerous cases missing from the evidence-backed submitted set \(\widehat{\mathcal{D}}\cap\mathcal{V}\).

Mitigation is evaluated by hidden revalidation after applying the submitted action plan \(\widehat{A}\). Let \(s_{\widehat{A}}(C)\) denote the post-action severity of case \(C\). Let \(\mathcal{M}\subseteq\mathcal{U}\) be the hidden mitigation check set, such as high-severity cases selected by the evaluator. The pre-action and post-action violation scores are
\begin{equation}
\mathrm{PreV}=
\frac{1}{|\mathcal{M}|}\sum_{C\in\mathcal{M}}s(C),
\qquad
\mathrm{PostV}=
\frac{1}{|\mathcal{M}|}\sum_{C\in\mathcal{M}}s_{\widehat{A}}(C).
\end{equation}
The violation reduction is
\begin{equation}
\mathrm{Red.}=
\frac{\mathrm{PreV}-\mathrm{PostV}}
{\max\{\epsilon,\mathrm{PreV}\}}.
\end{equation}
The action cost \(\mathrm{Cost}(\widehat{A})\) is task-specific and discourages unrealistic overuse of redispatch, curtailment, switching, or other controls.

The runner also reports workflow diagnostics computed from the evidence log. These include validation calls \(|\mathcal{V}|\), validation-budget use \(|\mathcal{V}|/B\), invalid tool calls, duplicate validation requests, explicit-submit rate, auto-finalization rate, schema repairs, type coercions, and wall-clock time when available. Results are reported as means and standard deviations over task instances or repeated trials, and reliability can also be summarized with pass@\(\ell\)-style metrics.

\subsection{Evaluation Protocol}
The evaluation protocol has three steps. First, the runner initializes a public task instance and exposes only the public tool API to the agent. Second, the agent interacts with the tools, while the runner records all calls, arguments, observations, validation results, and the final submission in the evidence log \(L\). Third, the runner sends the submitted \(R\) to the hidden evaluator for scoring. Scripted agents return \(R\) directly. LLM agents use the JSON adapter, which translates each model command into a public tool call; the runner executes the call, enforces budgets, updates \(L\), and stops when the agent calls \texttt{submit} or reaches the turn limit.

The protocol supports open, sealed, and stress regimes. In the open regime, users access the public case files and runner for development and reproducibility. The public repository used in this paper is an open-regime runner: it may compute the exhaustive oracle locally after constructing a case, but the agent still observes only public tool outputs. In the sealed regime, the same scoring interface can be hosted separately so oracle labels, test seeds, and evaluator code remain hidden from the agent. The stress regime is a reliability protocol layered on open or sealed runs: the same agent is evaluated across scenario variants, prompt perturbations, or repeated trials.

We use two scoring modes. In strict-submit scoring, failure to call \texttt{submit} gives an empty submitted set \(\widehat{\mathcal{D}}\), although validated cases still count toward \(R_{\mathrm{found}}\) and workflow diagnostics. In auto-finalized scoring, the runner may construct \(\widehat{\mathcal{D}}\) from validated cases, but this is marked by an auto-finalization diagnostic. Unsupported claims, duplicate validations, schema deviations, type mismatches, missing reports, and mitigation outcomes are computed from \(L\) and hidden validation, so the core score does not require an LLM judge.

\section{Case Study and Results}
\subsection{Pilot Environment}
We instantiate PowerAgentBench-SS with a DC thermal $N$-\(2\) contingency-search pilot on the IEEE 39-bus case. The hidden evaluation set \(\mathcal{U}\) is the set of all two-branch outage cases. The agent's goal is to identify high-severity cases \(C\in\mathcal{U}\), submit a ranked set \(\widehat{\mathcal{D}}\), support that set with public validation evidence \(\mathcal{V}\), and, when mitigation is attempted, reduce hidden post-action severity through bounded preventive redispatch.

Each test instance is one of 8 deterministic operating-point variants produced by applying fixed-seed load and dispatch perturbations to the same topology and equipment data. The candidate branch set contains 46 branches, so \(|\mathcal{U}|=\binom{46}{2}=1035\) $N$-\(2\) outage cases. The report size is \(m=20\), so \(\mathcal{O}_{m}=\mathcal{O}_{20}\). The dangerous set \(\mathcal{D}\) is the empirical top 5\% of \(\mathcal{U}\) by hidden pre-action severity, which gives 52 dangerous cases per instance. The mitigation check set is \(\mathcal{M}=\mathcal{O}_{20}\).

The severity function \(s(C)\) is a DC thermal-security proxy rather than a full AC security score. It is computed as the sum of normalized emergency-limit overloads plus an islanding penalty. The pilot does not model voltage constraints, reactive power limits, generator capability curves, transformer tap controls, or contingency-dependent emergency ratings. These modeling choices make the pilot reproducible and fast, but they should not be interpreted as a replacement for AC security analysis.

The validation budget is \(B=80\), so a budgeted agent may validate only 80 of the 1035 cases. Agents may submit 20 ranked contingencies. The hidden evaluator still computes the exhaustive oracle over \(\mathcal{U}\) for scoring. The public tool server exposes six tools. \texttt{case\_summary} returns network size, base-case severity, candidate count, budget, and report size. \texttt{rank\_base\_loading} ranks outage pairs by base-flow stress on the removed branches. \texttt{rank\_lodf} ranks outage pairs using a Line Outage Distribution Factors (LODF)-style approximate post-outage stress screen. These two ranking tools are prioritization aids and do not create validation evidence. \texttt{validate} performs exact public DC validation for selected cases, records evidence in \(\mathcal{V}\), and consumes budget. \texttt{redispatch} applies bounded preventive redispatch on a selected focus set. \texttt{submit} records the final ranked contingencies, mitigation, and diagnosis.

\subsection{Compared Agents}
We compare scripted baselines with LLM agents under the same public tools, budget, and hidden scoring. The scripted baselines isolate unsupported reporting, random search, topology-only screening, base-case stress, LODF-style physics screening, hybrid screening, and mitigation. \emph{No-validation} submits the top-ranked cases from the base-loading screen without calling \texttt{validate}, so \(\mathcal{V}=\emptyset\). \emph{Random} uniformly samples cases from \(\mathcal{U}\). \emph{Topology} ranks outage pairs by the degrees of their endpoint buses, where higher-degree buses have more incident branches \cite{kocc2014structural}. \emph{Base-loading} ranks pairs whose removed branches are highly loaded in the base case, a transparent stress heuristic related to contingency filtering \cite{capitanescu2011scopf,kaplunovich2016fast}. \emph{LODF-screen} uses an LODF-style approximate screen, which is common for fast contingency screening and $N$-\(x\) search-space reduction \cite{narimani2021generalized}. \emph{Hybrid-tool} combines top-ranked cases from Base-loading, LODF-screen, and Topology, removes duplicates, and validates the resulting prioritized list. \emph{Hybrid+redispatch} adds bounded preventive redispatch to test the discovery-to-mitigation transition. The LLM agents \emph{Qwen3.5}, \emph{Mistral-Nemo-12B}, and \emph{Command-R-35B} are served through Ollama and wrapped by the JSON-command adapter, while \emph{GPT-5.5} uses the OpenAI API. All LLM agents use the same public prompt, tool schema, strict-submit scoring, maximum turn limit, and budget. \emph{Exhaustive search} validates all 1035 cases and is a non-budgeted discovery reference.

\subsection{Results and Analysis}
Table~\ref{tab:results} compares discovery, evidence quality, and mitigation across budgeted scripted agents, strict-submit LLM agents, and the non-budgeted exhaustive-search reference. The table is designed to answer three questions at once: whether an agent recovers many oracle top-20 cases, whether its submitted cases are supported by validation evidence, and whether any mitigation reduces hidden post-action severity. Entries are reported as mean\(\pm\)standard deviation. Deterministic scripted agents and LLM agents are summarized over the 8 operating-point variants. Random is summarized over 50 random trials for each variant, i.e., 400 variant-trial runs.

\begin{table*}[t]
\centering
\caption{Strict-submit results for the IEEE 39-bus DC thermal $N$-\(2\) pilot.}
\label{tab:results}
\scriptsize
\resizebox{\textwidth}{!}{%
\begin{tabular}{@{}lcccccccc@{}}
\toprule
\textbf{Agent} & \(\boldsymbol{|\mathcal{V}|}\) & \(\boldsymbol{R_{\mathrm{evid}}}\) & \(\boldsymbol{R_{\mathrm{found}}}\) & \textbf{EvRate} & \textbf{Best} & \textbf{Regret} & \textbf{PostV} & \textbf{Red.} \\
\midrule
No-validation
& \(0.0{\pm}0.0\)
& \(0.000{\pm}0.000\)
& \(0.000{\pm}0.000\)
& \(0.000{\pm}0.000\)
& \(0.000{\pm}0.000\)
& \(17.942{\pm}0.055\)
& \(16.754{\pm}0.059\)
& \(0.000{\pm}0.000\) \\

Random
& \(80.0{\pm}0.0\)
& \(0.079{\pm}0.059\)
& \(0.079{\pm}0.059\)
& \(1.000{\pm}0.000\)
& \(0.932{\pm}0.047\)
& \(1.223{\pm}0.842\)
& \(16.754{\pm}0.059\)
& \(0.000{\pm}0.000\) \\

Topology
& \(80.0{\pm}0.0\)
& \(0.000{\pm}0.000\)
& \(0.000{\pm}0.000\)
& \(1.000{\pm}0.000\)
& \(0.350{\pm}0.021\)
& \(11.662{\pm}0.364\)
& \(16.754{\pm}0.059\)
& \(0.000{\pm}0.000\) \\

Base-loading
& \(80.0{\pm}0.0\)
& \(\mathbf{0.519{\pm}0.024}\)
& \(\mathbf{0.519{\pm}0.024}\)
& \(1.000{\pm}0.000\)
& \(\mathbf{1.000{\pm}0.000}\)
& \(\mathbf{0.000{\pm}0.000}\)
& \(16.754{\pm}0.059\)
& \(0.000{\pm}0.000\) \\

LODF-screen
& \(80.0{\pm}0.0\)
& \(0.081{\pm}0.024\)
& \(0.081{\pm}0.024\)
& \(1.000{\pm}0.000\)
& \(0.903{\pm}0.006\)
& \(1.743{\pm}0.108\)
& \(16.754{\pm}0.059\)
& \(0.000{\pm}0.000\) \\

Hybrid-tool
& \(80.0{\pm}0.0\)
& \(0.225{\pm}0.025\)
& \(0.225{\pm}0.025\)
& \(1.000{\pm}0.000\)
& \(\mathbf{1.000{\pm}0.000}\)
& \(\mathbf{0.000{\pm}0.000}\)
& \(16.754{\pm}0.059\)
& \(0.000{\pm}0.000\) \\

Hybrid+redispatch
& \(80.0{\pm}0.0\)
& \(0.225{\pm}0.025\)
& \(0.225{\pm}0.025\)
& \(1.000{\pm}0.000\)
& \(\mathbf{1.000{\pm}0.000}\)
& \(\mathbf{0.000{\pm}0.000}\)
& \(\mathbf{15.469{\pm}0.107}\)
& \(\mathbf{0.077{\pm}0.004}\) \\

Qwen3.5-Ollama
& \(80.0{\pm}0.0\)
& \(0.081{\pm}0.024\)
& \(0.081{\pm}0.024\)
& \(1.000{\pm}0.000\)
& \(0.903{\pm}0.006\)
& \(1.743{\pm}0.108\)
& \(16.754{\pm}0.059\)
& \(0.000{\pm}0.000\) \\

Mistral-Nemo-Ollama
& \(28.0{\pm}19.5\)
& \(0.013{\pm}0.033\)
& \(0.013{\pm}0.033\)
& \(1.000{\pm}0.000\)
& \(0.830{\pm}0.031\)
& \(3.054{\pm}0.567\)
& \(16.754{\pm}0.059\)
& \(0.000{\pm}0.000\) \\

Command-R-Ollama
& \(19.0{\pm}5.3\)
& \(0.019{\pm}0.050\)
& \(0.131{\pm}0.097\)
& \(0.125{\pm}0.331\)
& \(0.957{\pm}0.075\)
& \(0.772{\pm}1.336\)
& \(16.754{\pm}0.059\)
& \(0.000{\pm}0.000\) \\

GPT-5.5-OpenAI
& \(80.0{\pm}0.0\)
& \(0.300{\pm}0.214\)
& \(0.300{\pm}0.214\)
& \(1.000{\pm}0.000\)
& \(0.961{\pm}0.043\)
& \(0.691{\pm}0.770\)
& \(16.754{\pm}0.059\)
& \(0.000{\pm}0.000\) \\

\midrule
Exhaustive search
& \(1035.0{\pm}0.0\)
& \(1.000{\pm}0.000\)
& \(1.000{\pm}0.000\)
& \(1.000{\pm}0.000\)
& \(1.000{\pm}0.000\)
& \(0.000{\pm}0.000\)
& \(16.754{\pm}0.059\)
& \(0.000{\pm}0.000\) \\
\bottomrule
\end{tabular}%
}
\end{table*}

The first observation from Table~\ref{tab:results} is that broad oracle-top-20 recovery and single worst-case discovery are not the same capability. Base-loading is the strongest budgeted method for broad recovery, with \(R_{\mathrm{evid}}=0.519\pm0.024\) indicating that many high-severity $N$-\(2\) outages involve branches that are already stressed in the base case. Hybrid-tool and Hybrid+redispatch diversify the 80-call budget across base-loading, LODF-style, and topology screens. That diversification reduces broad recovery to \(R_{\mathrm{evid}}=0.225\pm0.025\), but it still finds the oracle worst case, giving Best \(=1.000\) and Regret \(=0.000\). Thus, Hybrid retains worst-case capture while sacrificing oracle-top-20 coverage relative to pure base loading. GPT-5.5 is the strongest LLM on broad recovery, with \(R_{\mathrm{evid}}=0.300\pm0.214\), but it still remains below the simple Base-loading screen. Random has low \(R_{\mathrm{evid}}\), as expected from sampling only 80 of 1035 cases, while Topology has zero oracle-top-20 recovery.

The second observation is that discovery and mitigation are separate outcomes. Base-loading, Hybrid-tool, and Hybrid+redispatch all identify the oracle worst case, but only Hybrid+redispatch changes the operating point. It reduces PostV from \(16.754\pm0.059\) to \(15.469\pm0.107\), corresponding to Red. \(=0.077\pm0.004\). This result is intentionally scored using separate quantities: \(\mathcal{V}\) records pre-action discovery evidence, while PostV and Red. are computed by hidden post-action revalidation on \(\mathcal{M}=\mathcal{O}_{20}\). Thus, mitigation does not overwrite the discovery evidence. No strict-submit LLM agent invokes mitigation in these runs and therefore their PostV values remain equal to PreV and their Red. values are zero.

The third observation is that strict-submit LLM scoring exposes workflow differences that would be hidden by answer-only evaluation. Qwen3.5 uses the full budget and has EvRate \(=1.000\), but its \(R_{\mathrm{evid}}\), Best, and Regret coincide with LODF-screen, indicating that its strategy reduces to validating the top candidates of the LODF-style screen. Mistral-Nemo submits explicitly and keeps all submitted cases evidence-backed, but it validates only \(28.0\pm19.5\) cases on average and therefore misses most of \(\mathcal{O}_{20}\). Command-R validates severe cases, with \(R_{\mathrm{found}}=0.131\pm0.097\) and Best \(=0.957\pm0.075\), but strict submission reduces \(R_{\mathrm{evid}}\) to \(0.019\pm0.050\). GPT-5.5 combines a clean workflow with stronger LLM search, achieving \(R_{\mathrm{evid}}=R_{\mathrm{found}}=0.300\pm0.214\), Best \(=0.961\pm0.043\), and Regret \(=0.691\pm0.770\). The relatively large Regret standard deviations for Command-R and GPT-5.5 are expected: both agents find the oracle worst case in some variants, giving Regret \(=0\), but miss it by a larger severity gap in other variants. The large \(R_{\mathrm{evid}}\) standard deviation of GPT-5.5 similarly reflects case-to-case variability across operating-point variants, which motivates the repeated-trial reliability reporting supported by the protocol. None of the LLM agents improves PostV under strict-submit scoring in this run.

\begin{table}[t]
\centering
\caption{Strict-submit workflow diagnostics for LLM agents.}
\label{tab:llm_diag}
\scriptsize
\begin{tabular}{@{}lccccc@{}}
\toprule
\textbf{Agent} & \(\boldsymbol{|\mathcal{V}|/B}\) & \textbf{Sub.} & \textbf{Dup.} & \textbf{Repair} & \textbf{Coerce} \\
\midrule
Qwen3.5
& \(1.00{\pm}0.00\)
& \(1.00{\pm}0.00\)
& \(0.0{\pm}0.0\)
& \(0.0{\pm}0.0\)
& \(0.0{\pm}0.0\) \\

Mistral-Nemo
& \(0.35{\pm}0.24\)
& \(1.00{\pm}0.00\)
& \(3.6{\pm}3.8\)
& \(0.0{\pm}0.0\)
& \(0.0{\pm}0.0\) \\

Command-R
& \(0.24{\pm}0.07\)
& \(0.12{\pm}0.33\)
& \(5.9{\pm}5.5\)
& \(0.0{\pm}0.0\)
& \(51.0{\pm}2.4\) \\

GPT-5.5
& \(1.00{\pm}0.00\)
& \(1.00{\pm}0.00\)
& \(0.0{\pm}0.0\)
& \(0.0{\pm}0.0\)
& \(0.0{\pm}0.0\) \\
\bottomrule
\end{tabular}
\end{table}

Table~\ref{tab:llm_diag} explains why the LLM rows in Table~\ref{tab:results} differ. Qwen3.5 and GPT-5.5 follow the intended workflow most closely: both use the full validation budget, avoid duplicate validation requests, and call \texttt{submit} in every instance. GPT-5.5 converts this clean workflow into stronger search performance, while Qwen3.5 mostly matches the scripted LODF-screen baseline. Mistral-Nemo also calls \texttt{submit} in every instance, but its budget use is only \(0.35\pm0.24\), and it makes \(3.6\pm3.8\) duplicate validation requests. Command-R has the opposite failure mode: it validates severe cases, but its explicit-submit rate is only \(0.12\pm0.33\), and it produces many type coercions because branch identifiers are often emitted in a noncanonical form. Under strict-submit scoring, those validated cases count toward \(R_{\mathrm{found}}\), but not toward \(R_{\mathrm{evid}}\) unless the agent also submits them in \(\widehat{\mathcal{D}}\).

\section{Conclusions}
PowerAgentBench-SS is a framework for evaluating LLM and tool-using agents on power system steady-state studies. It shifts the benchmark target from one-shot answers or isolated solver calls to an auditable engineering workflow: inspect, screen, validate, mitigate, and report. The $N$-\(2\) pilot on deterministic IEEE 39-bus operating-point variants demonstrates that the framework distinguishes unsupported answers, random tool use, heuristic screening, evidence-backed discovery, locally hosted and API-served LLM tool use, mitigation quality, and workflow-completion behavior under a fixed validation budget.

The pilot demonstrates the value of PowerAgentBench-SS as a reusable benchmark for future research and development of power system agents. By starting from a controlled DC thermal security-assessment setting, the benchmark provides a transparent environment in which agent behavior can be measured, compared, and diagnosed with physical grounding. The $N$-\(2\) pilot is not intended as a final ranking of foundation models, but as evidence that the benchmark can reveal meaningful differences in agentic behavior: how agents allocate validation budgets, whether they stop prematurely, how reliably they support claims with evidence, and whether they can translate contingency discovery into corrective action. These distinctions are essential for developing future agents that are not only accurate, but also auditable, budget-aware, and operationally useful.

Future development can extend the same protocol to AC power flow, richer control actions, larger test cases, SCOPF workflows, commercial simulator APIs, and planning studies. It can also support human-supervised settings where high-impact actions require approval. Across these extensions, the core principle remains unchanged: agents are evaluated by whether their tool-backed claims survive hidden engineering validation, not by how convincing their text appears.

\bibliographystyle{IEEEtran}
\bibliography{references}

\end{document}